\pdfoutput=1 
\documentclass[journal]{IEEEtran}

\usepackage{graphicx}
\usepackage{color}
\usepackage{natbib}
\usepackage{url}
\definecolor{darkgreen}{rgb}{.0,.3,.3}
\usepackage{hyperref}
\hypersetup{colorlinks,citecolor=darkgreen}
\usepackage{dblfloatfix}

\definecolor{mypurple}{rgb}{.4,.0,.6}
\newcommand{\preprint}[1]{\textcolor{mypurple}{\textbf{#1}}}

\newcommand{\vc}[1]{}
\newcommand{\gv}[1]{}

\begin{document}
%\title{Learning from failure - guidelines for improving machine learning in medical imaging}

\title{How I failed machine learning in medical imaging --
shortcomings and recommendations}

\author{
    \IEEEauthorblockN{Gaël Varoquaux\IEEEauthorrefmark{1}, Veronika Cheplygina\IEEEauthorrefmark{2}}\\
    \IEEEauthorblockA{\IEEEauthorrefmark{1}INRIA, France}\\
    \IEEEauthorblockA{\IEEEauthorrefmark{2}IT University of Copenhagen, Denmark}\\
}

\IEEEaftertitletext{%
\vspace*{-5ex}

{\hspace*{.05\linewidth}\fbox{\parbox{0.90\linewidth}{\preprint{This preprint
is a early version, and the final version has been published in NPJ
Digital Medicine under the title
\href{https://www.nature.com/articles/s41746-022-00592-y}{``\emph{Machine
learning for medical imaging: methodological failures and recommendations
for the future}''}. The new version contains updated text and figures,
please read and cite it, rather than the present document.%
}}}}%
\vspace*{2ex}
}

% make the title area
\maketitle
\IEEEpeerreviewmaketitle

\begin{abstract}
    Research in computer analysis of medical images bears many promises
to improve patients' health. However, a number of systematic challenges
are slowing down the progress of the field, from limitations of the data,
such as biases, to research incentives, such as optimizing for publication.
In this paper we review roadblocks to developing and assessing methods.
Building our analysis on evidence from the literature and data
challenges, we show that at every step, potential biases can creep in. On
a positive note, we also discuss on-going effort to counteract these problems. Finally we provide recommendations on how to further these address problems in the future. 
\end{abstract}

\section{Introduction}
Machine learning, the cornerstone of today's artificial intelligence (AI)
revolution, brings new promises to clinical practice with medical
images~\citep{litjens2017survey,cheplygina2019not,zhou2020review}. For
example, to diagnose various conditions from medical images, machine
learning has been shown to perform on par with medical
experts \citep[see][for a recent overview]{liu2019comparison}. Software
applications are starting to be certified for clinical
use~\citep{topol2019high,sendak2020path}. Machine learning may be the key
to realizing the vision of AI in medicine sketched several decades ago
\citep{schwartz1987artificial}.

The stakes are high, and there is a staggering amount of research on
machine learning for medical images. But this growth does not inherently lead to clinical
progress. The higher volume of research can be aligned with the
academic incentives rather than the needs of clinicians and patients. For
instance, there can be an oversupply of papers showing state-of-the-art performance on benchmark data, but no practical improvement for the clinical problem. 
On the topic of machine learning for COVID, 
Robert \emph{et al.}~\citep{roberts2021common} reviewed 62 published studies, but found none
with potential for clinical use.  

In this paper, we explore avenues to improve clinical impact of machine
learning in medical imaging. After sketching the situation,
documenting uneven progress in Section \ref{sec:sample_size}, we study a
number of failures frequent in medical imaging papers, at different steps of the ``publishing lifecycle'':
what data to use (Section \ref{sec:data}), 
what methods to use and how to evaluate them (Section
\ref{sec:evaluation}), and how to publish the results (Section \ref{sec:publishing}).
For reproducibility, data and code for our analyses are available on \url{https://github.com/GaelVaroquaux/ml_med_imaging_failures}.

In each section we first discuss the problems, supported with evidence
from previous research as well as our own analyses of recent papers. We then discuss a number of steps to
improve the situation, sometimes borrowed from related communities. We
hope that these ideas will help shape research practices that are even more
effective at addressing real-world medical challenges.

\section{It's not all about larger datasets}\label{sec:sample_size}
%Datasets have been getting larger, but, there is not always improvement! 
\begin{figure*}%[b!]
    \includegraphics[width=.31\linewidth]{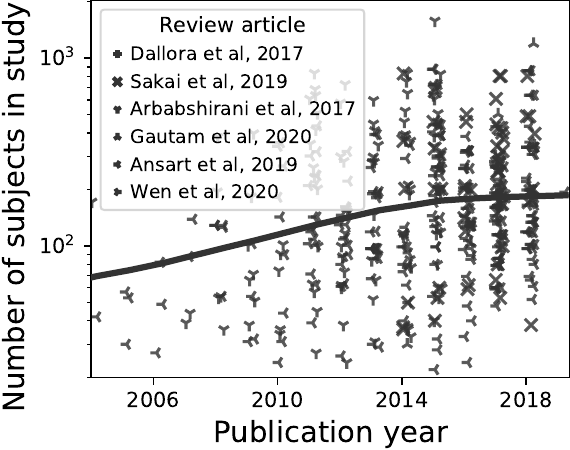}%
    \llap{\raisebox{1ex}{\sffamily\bfseries a.}\hspace*{.26\linewidth}}%
    \hfill%
    \includegraphics[width=.31\linewidth]{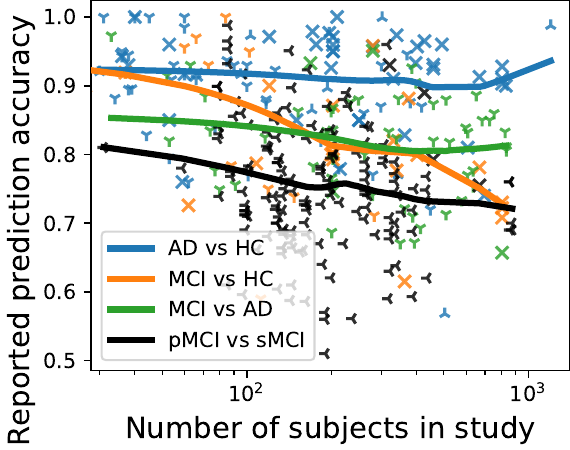}%
    \llap{\raisebox{1ex}{\sffamily\bfseries b.}\hspace*{.26\linewidth}}%
    \hfill%
    \includegraphics[width=.31\linewidth]{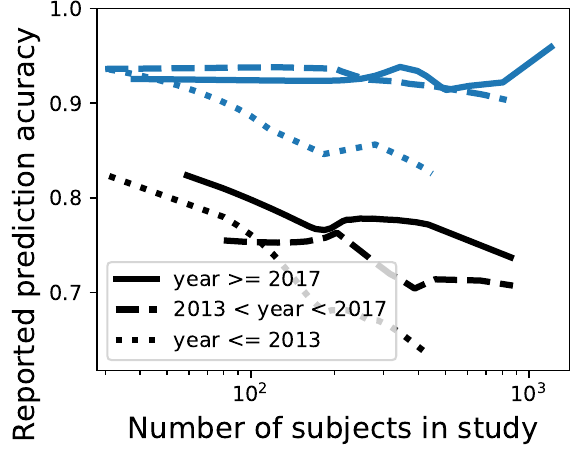}%
    \llap{\raisebox{1ex}{\sffamily\bfseries c.}\hspace*{.26\linewidth}}%

    \caption{\textbf{Bigger brain-imaging datasets are not enough for
better machine-learning diagnosis of Alzheimer's}. A meta-analysis across 6
review papers, covering more than 500 individual publications. The
machine-learning problem is typically formulated as distinguishing
various related clinical conditions, Alzheimer's Disease (AD), Healthy
Control (HC), and Mild Cognitive Impairment, which can signal prodromal
Alzheimer's. Distinguishing progressive mild cognitive impairment (pMCI)
from stable mild cognitive impairment (sMCI) is the most relevant
machine-learning task from the clinical standpoint.
\textbf{a.} Reported sample size as a function of the publication year of a study.
\textbf{b.} Reported prediction accuracy as a function of the number of subjects in a study.
\textbf{c.} Same plot distinguishing studies published in different years.
}
    \label{fig:sample_size}
\end{figure*}

The availability of large labeled datasets has enabled solving
difficult machine learning problems, such as natural image recognition in computer vision. As a
result, there is widespread hope that similar progress will happen in
medical applications: with large datasets,
algorithm research will eventually solve a clinical problem posed as discrimination task.
Few clinical questions come as well-posed discrimination tasks that can
be naturally framed as machine-learning tasks. But, even for these, larger datasets have often failed to lead to the progress hoped for.

One example is that of early diagnosis of Alzheimer's disease (AD), which
is a growing health burden due to the aging population. Early diagnosis
would open the door to early-stage interventions, which are most likely to be
effective. Substantial efforts have acquired large brain-imaging
cohorts of aging individuals at risk of developing AD,
on which early biomarkers can be developed using machine learning
\citep{mueller2005ways}. As a result, there have been steady increases in
the typical sample size of studies applying machine learning to develop
computer-aided diagnosis of AD, or its predecessor,
mild cognitive impairment. This growth is clearly visible in publications,
as on \autoref{fig:sample_size}a, a meta-analysis compiling
478 studies from 6 systematic reviews
\citep{dallora2017machine,arbabshirani2017single,liu2019comparison,sakai2019machine,wen2020convolutional,ansart2020predicting}.

However, the increase in data size did not come with better diagnostic
accuracy, in particular for the most clinically relevant question,
distinguishing pathological versus stable evolution for patients with
symptoms of prodromal Alzheimer's (\autoref{fig:sample_size}b). Rather,
studies with larger sample sizes tend to report worse prediction
accuracy. This is worrisome, as these larger studies are closer to
real-life settings. On the other hand, research efforts across time did lead to
improvements even on large, heterogeneous cohorts
(\autoref{fig:sample_size}c), as studies published later show
improvements for large sample sizes.

\section{Data, an imperfect window on the clinic}\label{sec:data}

\subsection{Datasets reflect an application only partly}
%OK, what if we have large datasets, are we done? 

Available datasets only partially reflect the clinical situation for a
particular medical condition, leading to dataset bias. This problem is
all the more important that the researcher may be unaware of this dataset
bias. Dataset bias occurs when the data used to build the decision model
(the training data), has a different distribution than the data
from the population on which it should be applied (the test data); for
example if the data were acquired with different scanners.  
As a result, algorithms which score high in
benchmarks can perform poorly in real world scenarios~\citep{zendel2017good}. 
In medical imaging, dataset bias has been demonstrated in
chest X-rays~\citep{pooch2019can,zech2018variable,larrazabal2020gender},
retinal imaging~\citep{tasdizen2018improving},
brain imaging~\citep{wachinger2021detect,ashraf2018learning},
histopathology~\citep{yu2018classify}, or
dermatology~\citep{abbasi2020risk}.
Such bias are revealed by training and testing a model across datasets from different sources, and observing a performance drop across sources. 

There are many potential sources of dataset bias in medical imaging,
introduced at different phases of the modeling
process~\citep{suresh2019framework}. First, a cohort may not appropriately represent the range of possible patients and symptoms, a bias sometimes called \emph{spectrum bias}
\citep{park2018methodologic}. 
A detrimental consequence is that model performance can be overestimated for different groups, for example between male and female
individuals~\citep{abbasi2020risk,larrazabal2020gender}. Yet medical imaging
publications seldom report the demographics of the
data.%

Imaging devices or procedures may lead to specific measurement biases.
A bias particularly harmful to clinically relevant automated diagnosis is when the data capture medical interventions. For instance,
on chest X-ray datasets, images for the
``pneumothorax'' condition sometimes show a chest drain, which is a
treatment for this condition, and which would not yet be present before
diagnosis \cite{oakden2020hidden}. Similar spurious correlations can appear in skin lesion images
due to markings placed by dermatologists next to the lesions~\citep{winkler2019association}.

Labeling errors can also introduce biases. Expert human annotators may
have systematic biases in the way they assign different
labels~\citep{joskowicz2019inter}, and it is seldom possible to
compensate with multiple annotators. Using automatic methods to
extract labels from patient reports can also lead to systematic errors
\citep{oakden2019exploring}. For example, a report on follow-up scan
often does not mention previously-known findings, which leads to
incorrect ``negative'' labels.

\subsection{Dataset availability distorts research}
%What to work on for this paper?
\begin{figure}[b]
    \centering
    \includegraphics[width=.9\columnwidth]{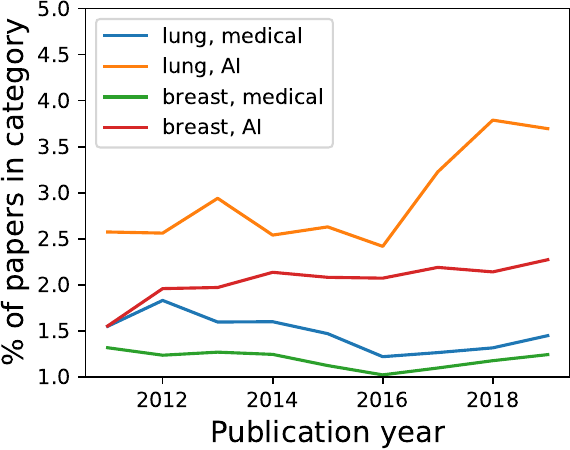}
 
    \caption{\textbf{Differences between popularity of applications.} We show the
percentage of papers on lung vs breast cancer, for medical oncology and
artificial intelligence. The percentages are relatively constant, except
lung cancer in AI, which shows an increase after 2016, year of the Kaggle
lung challenge.}
    \label{fig:publications}
\end{figure}

The availability of datasets can influence which applications are more studied. A striking example can be seen in oncology, with the
wide availability of various lung datasets on Kaggle or
\url{grand-challenge.org}, contrasted with (to our knowledge) only one
challenge focusing on mammograms. These data are associated with two
applications of medical imaging: detecting lung
nodules, and detecting breast tumors in radiological images.  The
prevalence of these topics in general medical oncology literature is
relatifvely constant across time, but in the AI literature 
lung imaging publications show a substantial increase in
2016 (Figure~\ref{fig:publications}, methodological details in Supp. Mat.
\ref{sec:popularity}). We suspect that the Kaggle lung challenges
published around that time contributed to this disproportional increase.

\subsection{Let us build awareness of data limitations}

Addressing such problems arising from the data requires critical thinking
about the choice of datasets, 
at the project level, i.e. which datasets to select for a study or a
challenge, and at a broader level, i.e. which datasets we work on as a community. 

At the project level, the choice of the dataset will influence the models
trained on the data, and the conclusions we can draw from the results. An
important step is using datasets from multiple sources, or creating
robust datasets from the start when
feasible~\citep{willemink2020preparing}. However, existing datasets can
still be critically evaluated for dataset
bias~\citep{rabanser2018failing}, hidden subgroups of
patients~\citep{oakden2020hidden}, mislabeled
instances~\citep{radsch2020radiologist}. 
A checklist for such
evaluation on computer vision datasets is presented in
\cite{zendel2017good}. When problems are discovered, relabeling a subset
of the data can be a worthwhile investment \citep{beyer2020done}. 

At the community level, we should foster understanding of
datasets limitations. Good documentation of datasets
should describe their characteristics and
data collection~\citep{datasheets2018}. Distributed models
should detail their limitations and the choices made to train them  (including the data)~\citep{mitchell2019model}.

Meta-analyses which look at evolution of dataset use in different areas are another way to reflect on current research efforts.
For example, a survey of crowdsourcing in medical
imaging~\citep{orting2019survey} shows a different distribution of
applications than surveys focusing on machine learning~\citep{litjens2017survey,cheplygina2019not}. Contrasting 
more clinically-oriented venues to more technical venues can reveal
opportunities for machine learning research.

%\gv{I think that there are solutions to measuring the robustness of an algorithm in the absence of a given bias, for instance \citep{chyzhyk2018controlling} (self-citation warning). \textbf{Assigned to Gaël}}

%-----------------------------------------
\section{Evaluations that miss the target}\label{sec:evaluation}

\subsection{Evaluation error is often larger than algorithmic improvements}%
\label{sec:evalerror}%
\begin{figure}[b!]
    \sffamily{\bfseries Evaluation noise in Kaggle
    competitions}\small\smallskip

    \begin{minipage}{.4\linewidth}
	Lung cancer

	Classification

	{\scriptsize Prize: \$1\,000\,000\\Test size: max 1K}
    \end{minipage}\hspace*{-.1\linewidth}%
    \begin{minipage}{.7\linewidth}
    \hspace*{-.06\linewidth}%
    \includegraphics[width=\linewidth]{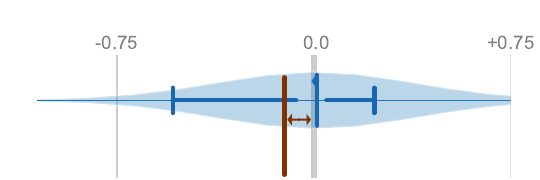}
    \end{minipage}%
    \vspace*{-.8ex}

    \begin{minipage}{.4\linewidth}
    Schizophrenia 
    
    Classification

	{\scriptsize Incentive: publications\\Test size: 120}\bigskip
    \end{minipage}\hspace*{-.1\linewidth}%
    \begin{minipage}{.7\linewidth}
    \hspace*{.04\linewidth}%
    \includegraphics[width=\linewidth]{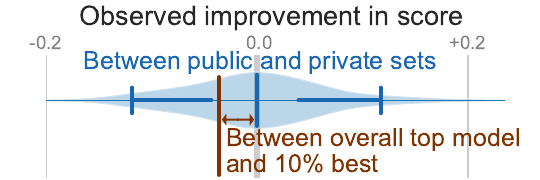}
    \end{minipage}
    \vspace*{-1.8ex}

    \begin{minipage}{.4\linewidth}
	Lung tumor

	Segmentation

	{\scriptsize Prize: \$30\,000\\Test size: max 6k}\smallskip
    \end{minipage}\hspace*{-.1\linewidth}%
    \begin{minipage}{.7\linewidth}
    \hspace*{.09\linewidth}%
    \includegraphics[width=\linewidth]{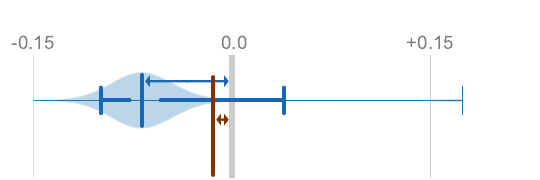}
    \end{minipage}%
    \vspace*{-1.8ex}

    \begin{minipage}{.4\linewidth}
	Nerve

	Segmentation

	{\scriptsize Prize: \$100\,000\\Test size 5.5K}\smallskip     \end{minipage}\hspace*{-.1\linewidth}%
    \begin{minipage}{.7\linewidth}
    \hspace*{-.01\linewidth}%
    \includegraphics[width=\linewidth]{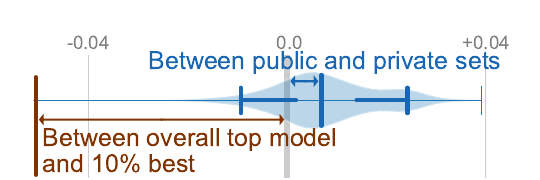}
    \end{minipage}%
    \caption{\textbf{Kaggle challenges: shifts from public to private
set compared to improvement across the top 10\% models} on 4
medical-imaging challenges with significant incentives. The blue violin
plot gives the distribution of differences between public and private
leaderboards (positive means that private leaderboard is better than public 
leaderboard). A systematic shift between public and private set indicates
overfittting or dataset bias. The width of this distribution gives the
intrinsic evaluation noise of the challenge.
The brown bar is $t_{10}$, the improvement between the top-most model (the winner) and
the 10\% best model. It is interesting to compare this improvement to the
shift and width in the difference between public and private set: if it
is smaller, the 10\% best models reached diminishing returns and did not
lead to a actual improvement on new data.
}
\label{fig:kaggle}
\end{figure}

Research on methods often focuses on outperforming other algorithms
on benchmark datasets. But too strong a focus on benchmark
performance can lead to
\emph{diminishing returns}, where increasingly large efforts achieve
smaller and smaller performance gains. Is such also visible in
the development of machine learning in medical imaging?

We studied performance improvements in four Kaggle medical-imaging
challenges, two on disease classification and two on image segmentation
(details in supp. mat. \ref{sec:challenges}).
We use the differences in algorithms performance between the public and
private leaderboards (two test sets used in the challenge) to quantify
the evaluation noise, in Figure~\ref{fig:kaggle}. We compare its distribution to the difference in
performance between the best algorithm, and the ``top 10\%'' algorithm.

Overall, three of the four challenges are in the diminishing returns category. For two challenges --schizophrenia and lung cancer diagnosis, with test set sizes below 1000-- the differences between performance on public and private test sets, are larger than improving 10\% on the leaderboard. For a third one, lung-tumor
segmentation, the performance on the private set is worse than on the
public set, revealing an overfit larger than the improvement. Only the nerve segmentation challenge displays substantial gains which are not
overfits.

\subsection{Improper evaluation procedures and leakage}%
\label{sec:leakage}%
Unbiased evaluation of model performance relies on training and testing the models with independent sets of data \citep{poldrack2020establishment}.
However incorrect implementations of this procedure can easily leak information,
leading to overoptimistic results. For example some studies classifying ADHD
based on brain imaging have engaged in circular analysis
\cite{pulini2019classification}, performing feature
selection on the full dataset, before cross-validation. Another
example of leakage arises when repeated measures of an individual are
split across train and test set, the algorithm then learning to recognize
the individual patient rather than markers of a condition \citep{saeb2017need}.

A related issue, yet more difficult to detect, is what we call
``overfitting by observer''. Even when cross-validation is carried out
for all steps of the method, overfitting may still occur by the
researcher adjusting the method to improve the
observed cross-validation performance \citep{skocik2016tried}. This can
explain some of the overfitting visible in challenges (Section
\ref{sec:evalerror}), though with challenges a private test set reveals
the overfitting, which is often not the case for published studies.

\subsection{Metrics that do not reflect what we want}
Evaluating models requires choosing a suitable metric. However, our
understanding of ``suitable'' may change over time. For example, the
image similarity metric which was widely used to evaluate image
registration algorithms, was later shown to be ineffective as scrambled
images could lead to high values~\citep{rohlfing2011image}. 

In medical image segmentation, \cite{maierhein2018rankings} review 150
challenges and show that the typical metrics used to rank algorithms are
sensitive to different variants of the same metric, casting doubt on the
objectivity of this algorithm ranking.

Important metrics may be missing from evaluation. Next to typical
classification metrics (sensitivity, specificity, area under the curve),
several authors argue for a calibration metric that compares the
predicted and observed probabilities~\citep{park2018methodologic,van2019calibration}. 

Finally, the metrics used may not be synonymous with practical
improvement~\citep{wagstaff2012machine,shankar2020evaluating}. For
example, typical metrics in
computer vision do not reflect important aspects of image
recognition, such as robustness to out-of-distribution examples \cite{shankar2020evaluating}.
Similarly, in medical imaging, improvements in traditional
metrics may not necessarily translate to different clinical
outcomes, e.g. robustness may be more important that an accurate
delineation in segmentation application.

\subsection{Incorrectly chosen baselines}%
\label{sec:baselines}%
%What should I compare my method to?

Developing new algorithms builds upon comparing these to baselines.
However, if these baselines are poorly chosen, the reported
improvement may be misleading. 

Baselines may not properly account for recent progress,
as revealed in applications of machine learning to
healthcare~\citep{bellamy2020evaluating}, but also other applications of machine learning \citep{oliver2018realistic,dacrema2019really,musgrave2020metric}. 

The opposite problem also occurs: forgetting simple approaches
effective for the problem at hand. For example, \cite{wen2020convolutional}
show that convolutional neural networks do not outperform support vector
machines for diagnosis of Alzheimer's disease from brain
imaging.

\subsection{Statistical significance not tested, or misunderstood}%
\label{sec:stat_test}%
%Now we got the performances on the test set, what can we conclude?

Experimental results are by nature noisy: results may
depend on which specific samples were used to train the models, the
random initializations, small differences in
hyper-parameters
\citep{bouthillier2019unreproducible,bouthillier2021accounting}. 
However, benchmarking predictive
models currently lacks well-adopted statistical good practices to
separate out this noise from generalizable findings.

A first, well-documented, source of brittleness arises from 
machine-learning experiments with too small sample
sizes~\citep{varoquaux2018cross,airola2009comparison}. Indeed, testing
predictive modeling require many samples, more than conventional
inferential studies, else the measured prediction accuracy may be a
distant estimation of real-life performance.
Sample sizes are growing, albeit slowly~\citep{szucs2020sample}. On a
positive note, a meta-analysis of public
vs private leaderboards on Kaggle \cite{roelofs2019meta} suggests that overfitting is less of an issue with ``large enough'' test
data (at least several thousands).  

Another challenge is that strong validation of a method requires it to be
robust to details of the data. Hence validation should go beyond
a single dataset, and rather strive for statistical
consensus across multiple datasets~\citep{demsar2006statistical}.
Yet, the corresponding statistical procedures require dozens of datasets to
establish significance and are seldom used in practice.
Rather, medical imaging research often reuses the same datasets across
studies, which raises the risk of finding an algorithm that performs well
by chance, in an implicit multiple comparison
problem \citep{thompson2020meta}.

But overall medical imaging research seldom analyzes how likely
empirical results are to be due to chance: only 6\% of segmentation
challenges surveyed \citep{maierhein2018winner}, and 15\% out of 410
popular computer science papers published by ACM used
a statistical test \citep{cockburn2020threats}.

However, null-hypothesis testing does not bring a full answer, as outlined
in~\cite{demsar2008appropriateness} by the very author of seminal
statistical testing work
\citep{demsar2006statistical}. Null-hypothesis tests are often misinterpreted
\citep{gigerenzer2018statistical}, with two notable challenges:
\emph{1)} the lack of statistically significant results
does not demonstrate the absence of effect, and \emph{2)} any trivial
effect can be significant given enough data
\citep{benavoli2016should,berrar2017confidence}.
For these reasons, \cite{bouthillier2019unreproducible} recommend to
replace traditional null-hypothesis testing by superiority testing, testing
that the improvement is above a given threshold.

%A last challenge is that publication incentives erode statistical control, as discussed in \autoref{sec:optimizing_publication}.

\subsection{Let us redefine evaluation}

\paragraph{Higher standards for benchmarking}

Good machine-learning benchmarks are difficult.
We compile below recognized best practices for medical machine learning evaluation
\citep{park2018methodologic,poldrack2020establishment,norgeot2020minimum}:
\begin{itemize}

\item Safeguarding from data leakage by separating out all test data from
the start, before any data transformation.

\item A documented way of selecting model hyper-parameters (including
architectural parameters for neural networks), without ever using data
from the test set.

\item Enough data in the test set to bring statistical power, at least
several hundreds samples, ideally thousands or more~\citep{willemink2020preparing}. 

\item Rich data to represent the diversity of patients and disease heterogeneity, ideally multi-institutional data including all relevant
patient demographics and disease state, with explicit inclusion criteria;
other cohorts with different recruitment go the extra mile to establish
external validity
\citep{steyerberg2016prediction,woo2017building}.

\item Strong baselines that reflect the state of the art of
machine-learning research, but also historical
solutions including clinical methodologies not necessarily relying on medical imaging.

\item A discussion the variability of the results due to
arbitrary choices (random seeds) and data sources with an eye on
statistical significance \cite{bouthillier2021accounting}.

\item Using different quantitative metrics to capture the different
aspects of the clinical problem and relating them to relevant clinical
performance metric.

\item Adding qualitative accounts and involving groups that will be most affected by the application in the metric design~\citep{thomas2020problem}.

\end{itemize}

\paragraph{More than beating the benchmark}

Even with proper validation and statistical
significance testing, measuring a tiny improvement on a benchmark is
seldom useful. Rather, one view is that, beyond rejecting a null,
a method should be accepted based on evidence that
it brings a sizable improvement upon the existing solutions. This type of
criteria is related to \emph{superiority
tests} used in clinical trials \citep{committee2001points,dagostino2003noninferiority,christensen2007methodology,schumi2011through}. For predictive modeling benchmarks, it amounts to comparing the observed improvement to
variation of the results due to arbitrary choices such as data sampling or random seeds
\citep{bouthillier2021accounting}.

Organizing blinded
challenges, with a hidden test set, mitigate the winner's curse. But to
bring progress, challenges should not to only focus on the winner. Instead,
more can be learned by comparing the competing methods and
analyzing the determinants of success, as well as failure cases.

%

%-----------------------------------------
\section{Publishing, distorted incentives}\label{sec:publishing}

\subsection{No incentive for clarity}
The publication process does not create incentives for clarity. Efforts
to impress may give rise to unnecessary ``mathiness'' of papers or suggestive
language (such as ``human-level performance'')
\cite{lipton2019troubling}. %Half-joking,  \cite{bearnensquash2010paper}
%argues that including equations increases the chance a paper is accepted
%to a prestigious conference. 
%
Important details may be omitted, from ablation
experiments showing what part of the method drives
improvements~\citep{lipton2019troubling}, to reporting how algorithms
were evaluated in a challenge~\citep{maierhein2018rankings}. This in turn
undermines reproducibility: being able to reproduce the exact
results or even draw the same conclusions~\citep{tatman2018practical,gundersen2018state}.

\subsection{Optimizing for publication}\label{sec:optimizing_publication}
As researchers our goal should be to solve scientific
problems. Yet, the reality of the culture we exist in can distort this
objective. Goodhart's law summarizes well the problem:
\emph{when
a measure becomes a target, it ceases to be a good measure}. As our
academic incentive system is based publications, it
erodes their scientific content via Goodhart's law.

Methods publication are selected for their novelty.
Yet, comparing 179 classifiers on 121 datasets shows no
statistically significant differences between the top
methods \citep{fernandez-delgado2014do}.
In order to sustain novelty,
researchers may be introducing unnecessary complexity into the methods,
that do not improve their prediction but rather
contribute to technical debt, making
systems harder to maintain and deploy~\citep{sculley2015hidden}.  

Another metric emphasized is obtaining ``state-of-the-art''
results, which leads to several of the evaluation problems outlined in
Section~\ref{sec:evaluation}.  The pressure to publish ``good'' results
can aggravate methodological loopholes~\citep{ioannidis2005most}, for
instance gaming the evaluation in machine
learning~\citep{teney2020value}. It is then all too appealing to find
after-the-fact theoretical justifications of positive yet fragile
empirical findings. This phenomenon, known as \emph{HARKing}
(hypothesizing after the results are known)~\citep{kerr1998harking}, has been documented in machine
learning~\citep{gencoglu2019hark} and computer science in
general~\citep{cockburn2020threats}.

Finally, the selection of publications creates the so-called ``file
drawer problem''~\citep{rosenthal1979file}: positive results, some due to
experimental flukes, are more likely to be published than corresponding
negative findings. For example,
in
410 most downloaded papers from the ACM, 97\% of the papers which used
significance testing had a finding with p-value of less than 0.05 \cite{cockburn2020threats}. It seems highly unlikely that only 3\% of the initial working hypotheses --even for impactful work-- turned out not confirmed.

\subsection{Let us improve our publication norms}

Fortunately there are various alleys to improve reporting and
transparency. For instance, the growing set of open datasets
could be leveraged for collaborative work beyond the
capacities of a single team~\citep{kellmeyer2017ethical}. 
The set of metrics studied could then be broadened, shifting the
publication focus away from a single-dimension benchmark. More metrics
can indeed help understanding a method's strengths and weaknesses,
exploring for instance calibration metrics
\citep{park2018methodologic,han2016develop,van2019calibration} or learning curves~\citep{richter2020sample}. Tutorials on the choice and estimation of metrics can be found in \citep{japkowicz2015performance,santafe2015dealing,pulini2019classification}. 

Method should be studied on more than prediction accuracy: 
reproducibility~\citep{gundersen2018state} carbon
footprint~\citep{henderson2020towards}, or a broad evaluation of
costs should be put in perspective with
the real-world patient outcomes, from a putative clinical use of the algorithms 
\citep{bowen2015increasing}. 

Preregistration or
registered reports can bring more robustness and trust: the motivation
and experimental setup of a paper are to be reviewed before empirical
results are available, and thus the paper is
be accepted before the experiments are run. Translating this idea to
machine learning faces the challenge that new data is seldom acquired in
a machine learning study, yet it would bring sizeable benefits~\citep{forde2019scientific,cockburn2020threats}. 
 
More generally, accelerating the progress in science calls for
accepting that some published findings are sometimes
wrong~\citep{firestein2015failure}. Popularizing different types of
publications may help, for example publishing negative
results~\citep{borji2018negative}, replication
studies~\citep{voets2018replication},
commentaries~\citep{wilkinson2020time} and retrospectives (such as the
recent NeurIPS Retrospectives workshop). Such initiatives should ideally
be led by more established academics, and welcoming of newcomers~\citep{whitaker2020bropenscience}.

%More discussion in papers! 
%documenting code~\citep{lee2018ten}

\section{Conclusions}

Despite great promises, the extensive research in medical applications of
machine learning seldom achieves a clinical impact.
Studying the academic literature and data-science challenges reveals
troubling trends: accuracy on diagnostic tasks progresses slower on
research cohorts that are closer to real-life settings; methods research
is often guided by dataset availability rather than clinical relevance;
many developments of model bring improvements smaller than the evaluation
errors.
We have surveyed challenges of clinical machine-learning research
that can explain these difficulties. 
The challenges start with the choice of datasets, plague model
evaluation, and are amplified by publication incentives.
Understanding these mechanisms enables us to suggest specific
strategies to improve the various steps of the research cycle, promoting
publications best practices
\cite{kakarmath2020best}. None of these strategies are
silver-bullet solutions. They rather require changing procedures, norms,
and goals. But implementing them will help fulfilling the promises of
machine-learning in
healthcare: better health outcomes for patients with less burden on the
care system.

\section*{Acknowledgements}

We would like to thank Alexandra Elbakyan for help with the literature review. We thank Pierre Dragicevic for providing feedback on early versions of this manuscript. 

% NOT included
% https://www.nature.com/articles/s41386-020-0767-z - from anger to failing
% https://arxiv.org/abs/2010.11008 - discussed, not relevant? 

\bibliographystyle{abbrvnat}
\bibliography{bib_refs_public}

\begin{thebibliography}{96}
\providecommand{\natexlab}[1]{#1}
\providecommand{\url}[1]{\texttt{#1}}
\expandafter\ifx\csname urlstyle\endcsname\relax
  \providecommand{\doi}[1]{doi: #1}\else
  \providecommand{\doi}{doi: \begingroup \urlstyle{rm}\Url}\fi

\bibitem[Abbasi-Sureshjani et~al.(2020)Abbasi-Sureshjani, Raumanns, Michels,
  Schouten, and Cheplygina]{abbasi2020risk}
S.~Abbasi-Sureshjani, R.~Raumanns, B.~E. Michels, G.~Schouten, and
  V.~Cheplygina.
\newblock Risk of training diagnostic algorithms on data with demographic bias.
\newblock In \emph{Interpretable and Annotation-Efficient Learning for Medical
  Image Computing}, pages 183--192. Springer, 2020.

\bibitem[Airola et~al.(2009)Airola, Pahikkala, Waegeman, De~Baets, and
  Salakoski]{airola2009comparison}
A.~Airola, T.~Pahikkala, W.~Waegeman, B.~De~Baets, and T.~Salakoski.
\newblock A comparison of {AUC} estimators in small-sample studies.
\newblock In \emph{Machine Learning in Systems Biology}, pages 3--13, 2009.

\bibitem[Ansart et~al.(2020)Ansart, Epelbaum, Bassignana, B{\^o}ne, Bottani,
  Cattai, Couronne, Faouzi, Koval, Louis, et~al.]{ansart2020predicting}
M.~Ansart, S.~Epelbaum, G.~Bassignana, A.~B{\^o}ne, S.~Bottani, T.~Cattai,
  R.~Couronne, J.~Faouzi, I.~Koval, M.~Louis, et~al.
\newblock Predicting the progression of mild cognitive impairment using machine
  learning: a systematic, quantitative and critical review.
\newblock \emph{Medical Image Analysis}, page 101848, 2020.

\bibitem[Arbabshirani et~al.(2017)Arbabshirani, Plis, Sui, and
  Calhoun]{arbabshirani2017single}
M.~R. Arbabshirani, S.~Plis, J.~Sui, and V.~D. Calhoun.
\newblock Single subject prediction of brain disorders in neuroimaging:
  Promises and pitfalls.
\newblock \emph{NeuroImage}, 145:\penalty0 137--165, 2017.

\bibitem[Ashraf et~al.(2018)Ashraf, Khan, Bhagwat, Chakravarty, and
  Taati]{ashraf2018learning}
A.~Ashraf, S.~Khan, N.~Bhagwat, M.~Chakravarty, and B.~Taati.
\newblock Learning to unlearn: building immunity to dataset bias in medical
  imaging studies.
\newblock In \emph{NeurIPS workshop on Machine Learning for Health (ML4H)}.
  2018.

\bibitem[Bellamy et~al.(2020)Bellamy, Celi, and Beam]{bellamy2020evaluating}
D.~Bellamy, L.~Celi, and A.~L. Beam.
\newblock Evaluating progress on machine learning for longitudinal electronic
  healthcare data.
\newblock \emph{arXiv preprint arXiv:2010.01149}, 2020.

\bibitem[Benavoli et~al.(2016)Benavoli, Corani, and
  Mangili]{benavoli2016should}
A.~Benavoli, G.~Corani, and F.~Mangili.
\newblock Should we really use post-hoc tests based on mean-ranks?
\newblock \emph{The Journal of Machine Learning Research}, 17\penalty0
  (1):\penalty0 152--161, 2016.

\bibitem[Berrar(2017)]{berrar2017confidence}
D.~Berrar.
\newblock Confidence curves: an alternative to null hypothesis significance
  testing for the comparison of classifiers.
\newblock \emph{Machine Learning}, 106\penalty0 (6):\penalty0 911--949, 2017.

\bibitem[Beyer et~al.(2020)Beyer, H{\'e}naff, Kolesnikov, Zhai, and
  Oord]{beyer2020done}
L.~Beyer, O.~J. H{\'e}naff, A.~Kolesnikov, X.~Zhai, and A.~v.~d. Oord.
\newblock Are we done with {ImageNet}?
\newblock \emph{arXiv preprint arXiv:2006.07159}, 2020.

\bibitem[Borji(2018)]{borji2018negative}
A.~Borji.
\newblock Negative results in computer vision: A perspective.
\newblock \emph{Image and Vision Computing}, 69:\penalty0 1--8, 2018.

\bibitem[Bouthillier et~al.(2019)Bouthillier, Laurent, and
  Vincent]{bouthillier2019unreproducible}
X.~Bouthillier, C.~Laurent, and P.~Vincent.
\newblock Unreproducible research is reproducible.
\newblock In \emph{International Conference on Machine Learning (ICML)}, pages
  725--734, 2019.

\bibitem[Bouthillier et~al.(2021)Bouthillier, Delaunay, Bronzi, Trofimov,
  Nichyporuk, Szeto, Mohammadi~Sepahvand, Raff, Madan, Voleti, Kahou,
  Michalski, Arbel, Pal, Varoquaux, and Vincent]{bouthillier2021accounting}
X.~Bouthillier, P.~Delaunay, M.~Bronzi, A.~Trofimov, B.~Nichyporuk, J.~Szeto,
  N.~Mohammadi~Sepahvand, E.~Raff, K.~Madan, V.~Voleti, S.~E. Kahou,
  V.~Michalski, T.~Arbel, C.~Pal, G.~Varoquaux, and P.~Vincent.
\newblock Accounting for variance in machine learning benchmarks.
\newblock In \emph{Machine Learning and Systems}, 2021.

\bibitem[Bowen and Casadevall(2015)]{bowen2015increasing}
A.~Bowen and A.~Casadevall.
\newblock Increasing disparities between resource inputs and outcomes, as
  measured by certain health deliverables, in biomedical research.
\newblock \emph{Proceedings of the National Academy of Sciences}, 112\penalty0
  (36):\penalty0 11335--11340, 2015.

\bibitem[Cheplygina et~al.(2019)Cheplygina, de~Bruijne, and
  Pluim]{cheplygina2019not}
V.~Cheplygina, M.~de~Bruijne, and J.~P.~W. Pluim.
\newblock Not-so-supervised: a survey of semi-supervised, multi-instance, and
  transfer learning in medical image analysis.
\newblock \emph{Medical Image Analysis}, 54:\penalty0 280--296, 2019.

\bibitem[Christensen(2007)]{christensen2007methodology}
E.~Christensen.
\newblock Methodology of superiority vs. equivalence trials and non-inferiority
  trials.
\newblock \emph{Journal of Hepatology}, 46\penalty0 (5):\penalty0 947--954,
  2007.

\bibitem[Cockburn et~al.(2020)Cockburn, Dragicevic, Besan{\c{c}}on, and
  Gutwin]{cockburn2020threats}
A.~Cockburn, P.~Dragicevic, L.~Besan{\c{c}}on, and C.~Gutwin.
\newblock Threats of a replication crisis in empirical computer science.
\newblock \emph{Communications of the ACM}, 63\penalty0 (8):\penalty0 70--79,
  2020.

\bibitem[Dacrema et~al.(2019)Dacrema, Cremonesi, and
  Jannach]{dacrema2019really}
M.~F. Dacrema, P.~Cremonesi, and D.~Jannach.
\newblock Are we really making much progress? a worrying analysis of recent
  neural recommendation approaches.
\newblock In \emph{ACM Conference on Recommender Systems}, pages 101--109,
  2019.

\bibitem[D'Agostino~Sr et~al.(2003)D'Agostino~Sr, Massaro, and
  Sullivan]{dagostino2003noninferiority}
R.~B. D'Agostino~Sr, J.~M. Massaro, and L.~M. Sullivan.
\newblock Non-inferiority trials: design concepts and issues--the encounters of
  academic consultants in statistics.
\newblock \emph{Statistics in Medicine}, 22\penalty0 (2):\penalty0 169--186,
  2003.

\bibitem[Dallora et~al.(2017)Dallora, Eivazzadeh, Mendes, Berglund, and
  Anderberg]{dallora2017machine}
A.~L. Dallora, S.~Eivazzadeh, E.~Mendes, J.~Berglund, and P.~Anderberg.
\newblock Machine learning and microsimulation techniques on the prognosis of
  dementia: A systematic literature review.
\newblock \emph{PLoS ONE}, 12\penalty0 (6):\penalty0 e0179804, 2017.

\bibitem[Dem{\v{s}}ar(2006)]{demsar2006statistical}
J.~Dem{\v{s}}ar.
\newblock Statistical comparisons of classifiers over multiple data sets.
\newblock \emph{The Journal of Machine Learning Research}, 7:\penalty0 1--30,
  2006.

\bibitem[Dem{\v{s}}ar(2008)]{demsar2008appropriateness}
J.~Dem{\v{s}}ar.
\newblock On the appropriateness of statistical tests in machine learning.
\newblock In \emph{ICML workshop on Evaluation Methods for Machine Learning},
  page~65, 2008.

\bibitem[Fernández-Delgado et~al.(2014)Fernández-Delgado, Cernadas, Barro,
  Amorim, and Amorim Fernández-Delgado]{fernandez-delgado2014do}
M.~Fernández-Delgado, E.~Cernadas, S.~Barro, D.~Amorim, and D.~Amorim
  Fernández-Delgado.
\newblock Do we need hundreds of classifiers to solve real world classification
  problems?
\newblock \emph{Journal of Machine Learning Research}, 15:\penalty0 3133--3181,
  2014.
\newblock ISSN 1532-4435.

\bibitem[Firestein(2015)]{firestein2015failure}
S.~Firestein.
\newblock \emph{Failure: Why science is so successful}.
\newblock Oxford University Press, 2015.

\bibitem[for the Evaluation~of Medicinal~Products(2001)]{committee2001points}
E.~A. for the Evaluation~of Medicinal~Products.
\newblock Points to consider on switching between superiority and
  non-inferiority.
\newblock \emph{British Journal of Clinical Pharmacology}, 52\penalty0
  (3):\penalty0 223--228, 2001.

\bibitem[Forde and Paganini(2019)]{forde2019scientific}
J.~Z. Forde and M.~Paganini.
\newblock The scientific method in the science of machine learning.
\newblock In \emph{ICLR workshop on Debugging Machine Learning Models}. 2019.

\bibitem[Gebru et~al.(2018)Gebru, Morgenstern, Vecchione, Vaughan, Wallach,
  III, and Crawford]{datasheets2018}
T.~Gebru, J.~Morgenstern, B.~Vecchione, J.~W. Vaughan, H.~M. Wallach, H.~D.
  III, and K.~Crawford.
\newblock Datasheets for datasets.
\newblock In \emph{Workshop on Fairness, Accountability, and Transparency in
  Machine Learning}. 2018.

\bibitem[Gencoglu et~al.(2019)Gencoglu, van Gils, Guldogan, Morikawa,
  S{\"u}zen, Gruber, Leinonen, and Huttunen]{gencoglu2019hark}
O.~Gencoglu, M.~van Gils, E.~Guldogan, C.~Morikawa, M.~S{\"u}zen, M.~Gruber,
  J.~Leinonen, and H.~Huttunen.
\newblock {HARK} side of deep learning--from grad student descent to automated
  machine learning.
\newblock \emph{arXiv preprint arXiv:1904.07633}, 2019.

\bibitem[Gigerenzer(2018)]{gigerenzer2018statistical}
G.~Gigerenzer.
\newblock Statistical rituals: The replication delusion and how we got there.
\newblock \emph{Advances in Methods and Practices in Psychological Science},
  1\penalty0 (2):\penalty0 198--218, 2018.

\bibitem[Gundersen and Kjensmo(2018)]{gundersen2018state}
O.~E. Gundersen and S.~Kjensmo.
\newblock State of the art: Reproducibility in artificial intelligence.
\newblock In \emph{AAAI Conference on Artificial Intelligence}, 2018.

\bibitem[Han et~al.(2016)Han, Song, and Choi]{han2016develop}
K.~Han, K.~Song, and B.~W. Choi.
\newblock How to develop, validate, and compare clinical prediction models
  involving radiological parameters: study design and statistical methods.
\newblock \emph{Korean Journal of Radiology}, 17\penalty0 (3):\penalty0
  339--350, 2016.

\bibitem[Henderson et~al.(2020)Henderson, Hu, Romoff, Brunskill, Jurafsky, and
  Pineau]{henderson2020towards}
P.~Henderson, J.~Hu, J.~Romoff, E.~Brunskill, D.~Jurafsky, and J.~Pineau.
\newblock Towards the systematic reporting of the energy and carbon footprints
  of machine learning.
\newblock \emph{Journal of Machine Learning Research}, 21\penalty0
  (248):\penalty0 1--43, 2020.

\bibitem[Hosseini et~al.(2020)Hosseini, Powell, Collins, Callahan-Flintoft,
  Jones, Bowman, and Wyble]{skocik2016tried}
M.~Hosseini, M.~Powell, J.~Collins, C.~Callahan-Flintoft, W.~Jones, H.~Bowman,
  and B.~Wyble.
\newblock I tried a bunch of things: The dangers of unexpected overfitting in
  classification of brain data.
\newblock \emph{Neuroscience \& Biobehavioral Reviews}, 2020.

\bibitem[Ioannidis(2005)]{ioannidis2005most}
J.~P.~A. Ioannidis.
\newblock Why most published research findings are false.
\newblock \emph{PLoS Medicine}, 2\penalty0 (8):\penalty0 e124, 2005.

\bibitem[Japkowicz and Shah(2015)]{japkowicz2015performance}
N.~Japkowicz and M.~Shah.
\newblock Performance evaluation in machine learning.
\newblock In \emph{Machine Learning in Radiation Oncology}, pages 41--56.
  Springer, 2015.

\bibitem[Joskowicz et~al.(2019)Joskowicz, Cohen, Caplan, and
  Sosna]{joskowicz2019inter}
L.~Joskowicz, D.~Cohen, N.~Caplan, and J.~Sosna.
\newblock Inter-observer variability of manual contour delineation of
  structures in {CT}.
\newblock \emph{European Radiology}, 29\penalty0 (3):\penalty0 1391--1399,
  2019.

\bibitem[Kakarmath et~al.(2020)Kakarmath, Esteva, Arnaout, Harvey, Kumar, Muse,
  Dong, Wedlund, and Kvedar]{kakarmath2020best}
S.~Kakarmath, A.~Esteva, R.~Arnaout, H.~Harvey, S.~Kumar, E.~Muse, F.~Dong,
  L.~Wedlund, and J.~Kvedar.
\newblock Best practices for authors of healthcare-related artificial
  intelligence manuscripts.
\newblock \emph{NPJ Digital Medicine}, 3:\penalty0 134--134, 2020.

\bibitem[Kellmeyer(2017)]{kellmeyer2017ethical}
P.~Kellmeyer.
\newblock Ethical and legal implications of the methodological crisis in
  neuroimaging.
\newblock \emph{Cambridge Quarterly of Healthcare Ethics}, 26\penalty0
  (4):\penalty0 530--554, 2017.

\bibitem[Kerr(1998)]{kerr1998harking}
N.~L. Kerr.
\newblock {HARKing}: hypothesizing after the results are known.
\newblock \emph{Personality and social psychology review}, 2\penalty0
  (3):\penalty0 196--217, 1998.

\bibitem[Larrazabal et~al.(2020)Larrazabal, Nieto, Peterson, Milone, and
  Ferrante]{larrazabal2020gender}
A.~J. Larrazabal, N.~Nieto, V.~Peterson, D.~H. Milone, and E.~Ferrante.
\newblock Gender imbalance in medical imaging datasets produces biased
  classifiers for computer-aided diagnosis.
\newblock \emph{Proceedings of the National Academy of Sciences}, 2020.

\bibitem[Lipton and Steinhardt(2019)]{lipton2019troubling}
Z.~C. Lipton and J.~Steinhardt.
\newblock Troubling trends in machine learning scholarship: some {ML} papers
  suffer from flaws that could mislead the public and stymie future research.
\newblock \emph{Queue}, 17\penalty0 (1):\penalty0 45--77, 2019.

\bibitem[Litjens et~al.(2017)Litjens, Kooi, Bejnordi, Setio, Ciompi,
  Ghafoorian, van~der Laak, Van~Ginneken, and S{\'a}nchez]{litjens2017survey}
G.~Litjens, T.~Kooi, B.~E. Bejnordi, A.~A.~A. Setio, F.~Ciompi, M.~Ghafoorian,
  J.~A. van~der Laak, B.~Van~Ginneken, and C.~I. S{\'a}nchez.
\newblock A survey on deep learning in medical image analysis.
\newblock \emph{Medical Image Analysis}, 42:\penalty0 60--88, 2017.

\bibitem[Liu et~al.(2019)Liu, Faes, Kale, Wagner, Fu, Bruynseels, Mahendiran,
  Moraes, Shamdas, Kern, et~al.]{liu2019comparison}
X.~Liu, L.~Faes, A.~U. Kale, S.~K. Wagner, D.~J. Fu, A.~Bruynseels,
  T.~Mahendiran, G.~Moraes, M.~Shamdas, C.~Kern, et~al.
\newblock A comparison of deep learning performance against health-care
  professionals in detecting diseases from medical imaging: a systematic review
  and meta-analysis.
\newblock \emph{The Lancet Digital Health}, 2019.

\bibitem[Maier-Hein et~al.(2018{\natexlab{a}})Maier-Hein, Eisenmann, Reinke,
  Onogur, Stankovic, Scholz, Arbel, Bogunovic, Bradley, Carass,
  et~al.]{maierhein2018rankings}
L.~Maier-Hein, M.~Eisenmann, A.~Reinke, S.~Onogur, M.~Stankovic, P.~Scholz,
  T.~Arbel, H.~Bogunovic, A.~P. Bradley, A.~Carass, et~al.
\newblock Why rankings of biomedical image analysis competitions should be
  interpreted with care.
\newblock \emph{Nature Communications}, 9\penalty0 (1):\penalty0 5217,
  2018{\natexlab{a}}.

\bibitem[Maier-Hein et~al.(2018{\natexlab{b}})Maier-Hein, Eisenmann, Reinke,
  Onogur, Stankovic, Scholz, Arbel, Bogunovic, Bradley, Carass,
  et~al.]{maierhein2018winner}
L.~Maier-Hein, M.~Eisenmann, A.~Reinke, S.~Onogur, M.~Stankovic, P.~Scholz,
  T.~Arbel, H.~Bogunovic, A.~P. Bradley, A.~Carass, et~al.
\newblock Is the winner really the best? a critical analysis of common research
  practice in biomedical image analysis competitions.
\newblock \emph{Nature Communications}, 2018{\natexlab{b}}.

\bibitem[Mitchell et~al.(2019)Mitchell, Wu, Zaldivar, Barnes, Vasserman,
  Hutchinson, Spitzer, Raji, and Gebru]{mitchell2019model}
M.~Mitchell, S.~Wu, A.~Zaldivar, P.~Barnes, L.~Vasserman, B.~Hutchinson,
  E.~Spitzer, I.~D. Raji, and T.~Gebru.
\newblock Model cards for model reporting.
\newblock In \emph{Fairness, Accountability, and Transparency (FAccT)}, pages
  220--229. ACM, 2019.

\bibitem[Mori and Taylor(2018)]{mori2018dimensions}
A.~Mori and M.~Taylor.
\newblock Dimensions metrics {API} reference \& getting started.
\newblock \emph{Digital Science \& Research solutions}, 2018.

\bibitem[Mueller et~al.(2005)Mueller, Weiner, Thal, Petersen, Jack, Jagust,
  Trojanowski, Toga, and Beckett]{mueller2005ways}
S.~G. Mueller, M.~W. Weiner, L.~J. Thal, R.~C. Petersen, C.~R. Jack, W.~Jagust,
  J.~Q. Trojanowski, A.~W. Toga, and L.~Beckett.
\newblock Ways toward an early diagnosis in {Alzheimer’s} disease: the
  {Alzheimer’s Disease Neuroimaging Initiative (ADNI)}.
\newblock \emph{Alzheimer's \& Dementia}, 1\penalty0 (1):\penalty0 55--66,
  2005.

\bibitem[Musgrave et~al.(2020)Musgrave, Belongie, and Lim]{musgrave2020metric}
K.~Musgrave, S.~Belongie, and S.-N. Lim.
\newblock A metric learning reality check.
\newblock In \emph{European Conference on Computer Vision}, pages 681--699.
  Springer, 2020.

\bibitem[Norgeot et~al.(2020)Norgeot, Quer, Beaulieu-Jones, Torkamani, Dias,
  Gianfrancesco, Arnaout, Kohane, Saria, Topol, et~al.]{norgeot2020minimum}
B.~Norgeot, G.~Quer, B.~K. Beaulieu-Jones, A.~Torkamani, R.~Dias,
  M.~Gianfrancesco, R.~Arnaout, I.~S. Kohane, S.~Saria, E.~Topol, et~al.
\newblock Minimum information about clinical artificial intelligence modeling:
  the {MI-CLAIM} checklist.
\newblock \emph{Nature Medicine}, 26\penalty0 (9):\penalty0 1320--1324, 2020.

\bibitem[Oakden-Rayner(2020)]{oakden2019exploring}
L.~Oakden-Rayner.
\newblock Exploring large-scale public medical image datasets.
\newblock \emph{Academic Radiology}, 27\penalty0 (1):\penalty0 106--112, 2020.

\bibitem[Oakden-Rayner et~al.(2020)Oakden-Rayner, Dunnmon, Carneiro, and
  R{\'e}]{oakden2020hidden}
L.~Oakden-Rayner, J.~Dunnmon, G.~Carneiro, and C.~R{\'e}.
\newblock Hidden stratification causes clinically meaningful failures in
  machine learning for medical imaging.
\newblock In \emph{ACM Conference on Health, Inference, and Learning}, pages
  151--159, 2020.

\bibitem[Oliver et~al.(2018)Oliver, Odena, Raffel, Cubuk, and
  Goodfellow]{oliver2018realistic}
A.~Oliver, A.~Odena, C.~Raffel, E.~D. Cubuk, and I.~J. Goodfellow.
\newblock Realistic evaluation of semi-supervised learning algorithms.
\newblock In \emph{Neural Information Processing Systems (NeurIPS)}, 2018.

\bibitem[{\O}rting et~al.(2020){\O}rting, Doyle, van Hilten, Hirth, Inel,
  Madan, Mavridis, Spiers, and Cheplygina]{orting2019survey}
S.~N. {\O}rting, A.~Doyle, A.~van Hilten, M.~Hirth, O.~Inel, C.~R. Madan,
  P.~Mavridis, H.~Spiers, and V.~Cheplygina.
\newblock A survey of crowdsourcing in medical image analysis.
\newblock \emph{Human Computation}, 7:\penalty0 1--26, 2020.

\bibitem[Park and Han(2018)]{park2018methodologic}
S.~H. Park and K.~Han.
\newblock Methodologic guide for evaluating clinical performance and effect of
  artificial intelligence technology for medical diagnosis and prediction.
\newblock \emph{Radiology}, 286\penalty0 (3):\penalty0 800--809, 2018.

\bibitem[Poldrack et~al.(2020)Poldrack, Huckins, and
  Varoquaux]{poldrack2020establishment}
R.~A. Poldrack, G.~Huckins, and G.~Varoquaux.
\newblock Establishment of best practices for evidence for prediction: a
  review.
\newblock \emph{JAMA Psychiatry}, 77\penalty0 (5):\penalty0 534--540, 2020.

\bibitem[Pooch et~al.(2019)Pooch, Ballester, and Barros]{pooch2019can}
E.~H. Pooch, P.~L. Ballester, and R.~C. Barros.
\newblock Can we trust deep learning models diagnosis? the impact of domain
  shift in chest radiograph classification.
\newblock In \emph{MICCAI workshop on Thoracic Image Analysis}. Springer, 2019.

\bibitem[Pulini et~al.(2019)Pulini, Kerr, Loo, and
  Lenartowicz]{pulini2019classification}
A.~A. Pulini, W.~T. Kerr, S.~K. Loo, and A.~Lenartowicz.
\newblock Classification accuracy of neuroimaging biomarkers in
  attention-deficit/hyperactivity disorder: Effects of sample size and circular
  analysis.
\newblock \emph{Biological Psychiatry: Cognitive Neuroscience and
  Neuroimaging}, 4\penalty0 (2):\penalty0 108--120, 2019.

\bibitem[Rabanser et~al.(2018)Rabanser, G{\"u}nnemann, and
  Lipton]{rabanser2018failing}
S.~Rabanser, S.~G{\"u}nnemann, and Z.~C. Lipton.
\newblock Failing loudly: an empirical study of methods for detecting dataset
  shift.
\newblock In \emph{Neural Information Processing Systems (NeurIPS)}. 2018.

\bibitem[R{\"a}dsch et~al.(2020)R{\"a}dsch, Eckhardt, Leiser, Pandl, Thiebes,
  and Sunyaev]{radsch2020radiologist}
T.~R{\"a}dsch, S.~Eckhardt, F.~Leiser, K.~D. Pandl, S.~Thiebes, and A.~Sunyaev.
\newblock What your radiologist might be missing: using machine learning to
  identify mislabeled instances of {X-ray} images.
\newblock In \emph{Hawaii International Conference on System Sciences (HICSS)}.
  2020.

\bibitem[Richter and Khoshgoftaar(2020)]{richter2020sample}
A.~N. Richter and T.~M. Khoshgoftaar.
\newblock Sample size determination for biomedical big data with limited
  labels.
\newblock \emph{Network Modeling Analysis in Health Informatics and
  Bioinformatics}, 9\penalty0 (1):\penalty0 12, 2020.

\bibitem[Roberts et~al.(2021)Roberts, Driggs, Thorpe, Gilbey, Yeung, Ursprung,
  Aviles-Rivero, Etmann, McCague, Beer, et~al.]{roberts2021common}
M.~Roberts, D.~Driggs, M.~Thorpe, J.~Gilbey, M.~Yeung, S.~Ursprung, A.~I.
  Aviles-Rivero, C.~Etmann, C.~McCague, L.~Beer, et~al.
\newblock Common pitfalls and recommendations for using machine learning to
  detect and prognosticate for {COVID-19} using chest radiographs and {CT}
  scans.
\newblock \emph{Nature Machine Intelligence}, 3\penalty0 (3):\penalty0
  199--217, 2021.

\bibitem[Roelofs et~al.(2019)Roelofs, Shankar, Recht, Fridovich-Keil, Hardt,
  Miller, and Schmidt]{roelofs2019meta}
R.~Roelofs, V.~Shankar, B.~Recht, S.~Fridovich-Keil, M.~Hardt, J.~Miller, and
  L.~Schmidt.
\newblock A meta-analysis of overfitting in machine learning.
\newblock In \emph{Neural Information Processing Systems (NeurIPS)}, pages
  9179--9189, 2019.

\bibitem[Rohlfing(2011)]{rohlfing2011image}
T.~Rohlfing.
\newblock Image similarity and tissue overlaps as surrogates for image
  registration accuracy: widely used but unreliable.
\newblock \emph{IEEE Transactions on Medical Imaging}, 31\penalty0
  (2):\penalty0 153--163, 2011.

\bibitem[Rosenthal(1979)]{rosenthal1979file}
R.~Rosenthal.
\newblock The file drawer problem and tolerance for null results.
\newblock \emph{Psychological Bulletin}, 86\penalty0 (3):\penalty0 638, 1979.

\bibitem[Saeb et~al.(2017)Saeb, Lonini, Jayaraman, Mohr, and
  Kording]{saeb2017need}
S.~Saeb, L.~Lonini, A.~Jayaraman, D.~C. Mohr, and K.~P. Kording.
\newblock The need to approximate the use-case in clinical machine learning.
\newblock \emph{Gigascience}, 6\penalty0 (5):\penalty0 gix019, 2017.

\bibitem[Sakai and Yamada(2019)]{sakai2019machine}
K.~Sakai and K.~Yamada.
\newblock Machine learning studies on major brain diseases: 5-year trends of
  2014--2018.
\newblock \emph{Japanese Journal of Radiology}, 37\penalty0 (1):\penalty0
  34--72, 2019.

\bibitem[Santafe et~al.(2015)Santafe, Inza, and Lozano]{santafe2015dealing}
G.~Santafe, I.~Inza, and J.~A. Lozano.
\newblock Dealing with the evaluation of supervised classification algorithms.
\newblock \emph{Artificial Intelligence Review}, 44\penalty0 (4):\penalty0
  467--508, 2015.

\bibitem[Schumi and Wittes(2011)]{schumi2011through}
J.~Schumi and J.~T. Wittes.
\newblock Through the looking glass: understanding non-inferiority.
\newblock \emph{Trials}, 12\penalty0 (1):\penalty0 1--12, 2011.

\bibitem[Schwartz et~al.(1987)Schwartz, Patil, and
  Szolovits]{schwartz1987artificial}
W.~B. Schwartz, R.~S. Patil, and P.~Szolovits.
\newblock Artificial intelligence in medicine, 1987.

\bibitem[Sculley et~al.(2015)Sculley, Holt, Golovin, Davydov, Phillips, Ebner,
  Chaudhary, Young, Crespo, and Dennison]{sculley2015hidden}
D.~Sculley, G.~Holt, D.~Golovin, E.~Davydov, T.~Phillips, D.~Ebner,
  V.~Chaudhary, M.~Young, J.-F. Crespo, and D.~Dennison.
\newblock Hidden technical debt in machine learning systems.
\newblock In \emph{Neural Information Processing Systems (NeurIPS)}, pages
  2503--2511, 2015.

\bibitem[Sendak et~al.(2020)Sendak, D’Arcy, Kashyap, Gao, Nichols, Corey,
  Ratliff, and Balu]{sendak2020path}
M.~P. Sendak, J.~D’Arcy, S.~Kashyap, M.~Gao, M.~Nichols, K.~Corey,
  W.~Ratliff, and S.~Balu.
\newblock A path for translation of machine learning products into healthcare
  delivery.
\newblock \emph{European Medical Journal Innovations}, 10:\penalty0 19--00172,
  2020.

\bibitem[Shankar et~al.(2020)Shankar, Roelofs, Mania, Fang, Recht, and
  Schmidt]{shankar2020evaluating}
V.~Shankar, R.~Roelofs, H.~Mania, A.~Fang, B.~Recht, and L.~Schmidt.
\newblock Evaluating machine accuracy on imagenet.
\newblock In \emph{International Conference on Machine Learning (ICML)}, 2020.

\bibitem[Steyerberg and Harrell(2016)]{steyerberg2016prediction}
E.~W. Steyerberg and F.~E. Harrell.
\newblock Prediction models need appropriate internal, internal--external, and
  external validation.
\newblock \emph{Journal of Clinical Epidemiology}, 69:\penalty0 245--247, 2016.

\bibitem[Suresh and Guttag(2019)]{suresh2019framework}
H.~Suresh and J.~V. Guttag.
\newblock A framework for understanding unintended consequences of machine
  learning.
\newblock \emph{arXiv preprint arXiv:1901.10002}, 2019.

\bibitem[Szucs and Ioannidis(2020)]{szucs2020sample}
D.~Szucs and J.~P. Ioannidis.
\newblock Sample size evolution in neuroimaging research: an evaluation of
  highly-cited studies (1990-2012) and of latest practices (2017-2018) in
  high-impact journals.
\newblock \emph{NeuroImage}, page 117164, 2020.

\bibitem[Tasdizen et~al.(2018)Tasdizen, Sajjadi, Javanmardi, and
  Ramesh]{tasdizen2018improving}
T.~Tasdizen, M.~Sajjadi, M.~Javanmardi, and N.~Ramesh.
\newblock Improving the robustness of convolutional networks to appearance
  variability in biomedical images.
\newblock In \emph{International Symposium on Biomedical Imaging (ISBI)}, pages
  549--553. IEEE, 2018.

\bibitem[Tatman et~al.(2018)Tatman, VanderPlas, and Dane]{tatman2018practical}
R.~Tatman, J.~VanderPlas, and S.~Dane.
\newblock A practical taxonomy of reproducibility for machine learning
  research.
\newblock In \emph{ICML workshop on Reproducibility in Machine Learning}, 2018.

\bibitem[Teney et~al.(2020)Teney, Kafle, Shrestha, Abbasnejad, Kanan, and
  Hengel]{teney2020value}
D.~Teney, K.~Kafle, R.~Shrestha, E.~Abbasnejad, C.~Kanan, and A.~v.~d. Hengel.
\newblock On the value of out-of-distribution testing: an example of
  {Goodhart's Law}.
\newblock In \emph{Neural Information Processing Systems (NeurIPS)}, 2020.

\bibitem[Thomas and Uminsky(2020)]{thomas2020problem}
R.~Thomas and D.~Uminsky.
\newblock The problem with metrics is a fundamental problem for {AI}.
\newblock \emph{arXiv preprint arXiv:2002.08512}, 2020.

\bibitem[Thompson et~al.(2020)Thompson, Wright, Bissett, and
  Poldrack]{thompson2020meta}
W.~H. Thompson, J.~Wright, P.~G. Bissett, and R.~A. Poldrack.
\newblock Meta-research: Dataset decay and the problem of sequential analyses
  on open datasets.
\newblock \emph{eLife}, 9:\penalty0 e53498, 2020.

\bibitem[Topol(2019)]{topol2019high}
E.~J. Topol.
\newblock High-performance medicine: the convergence of human and artificial
  intelligence.
\newblock \emph{Nature Medicine}, 25\penalty0 (1):\penalty0 44--56, 2019.

\bibitem[Van~Calster et~al.(2019)Van~Calster, McLernon, Van~Smeden, Wynants,
  and Steyerberg]{van2019calibration}
B.~Van~Calster, D.~J. McLernon, M.~Van~Smeden, L.~Wynants, and E.~W.
  Steyerberg.
\newblock Calibration: the {Achilles} heel of predictive analytics.
\newblock \emph{BMC Medicine}, 17\penalty0 (1):\penalty0 1--7, 2019.

\bibitem[Varoquaux(2018)]{varoquaux2018cross}
G.~Varoquaux.
\newblock Cross-validation failure: small sample sizes lead to large error
  bars.
\newblock \emph{NeuroImage}, 180:\penalty0 68--77, 2018.

\bibitem[Voets et~al.(2018)Voets, M{\o}llersen, and
  Bongo]{voets2018replication}
M.~Voets, K.~M{\o}llersen, and L.~A. Bongo.
\newblock Replication study: Development and validation of deep learning
  algorithm for detection of diabetic retinopathy in retinal fundus
  photographs.
\newblock \emph{arXiv preprint arXiv:1803.04337}, 2018.

\bibitem[Wachinger et~al.(2021)Wachinger, Rieckmann, P{\"o}lsterl, Initiative,
  et~al.]{wachinger2021detect}
C.~Wachinger, A.~Rieckmann, S.~P{\"o}lsterl, A.~D.~N. Initiative, et~al.
\newblock Detect and correct bias in multi-site neuroimaging datasets.
\newblock \emph{Medical Image Analysis}, 67:\penalty0 101879, 2021.

\bibitem[Wagstaff(2012)]{wagstaff2012machine}
K.~L. Wagstaff.
\newblock Machine learning that matters.
\newblock In \emph{International Conference on Machine Learning (ICML)}, pages
  529--536, 2012.

\bibitem[Wen et~al.(2020)Wen, Thibeau-Sutre, Diaz-Melo, Samper-Gonz{\'a}lez,
  Routier, Bottani, Dormont, Durrleman, Burgos, Colliot,
  et~al.]{wen2020convolutional}
J.~Wen, E.~Thibeau-Sutre, M.~Diaz-Melo, J.~Samper-Gonz{\'a}lez, A.~Routier,
  S.~Bottani, D.~Dormont, S.~Durrleman, N.~Burgos, O.~Colliot, et~al.
\newblock Convolutional neural networks for classification of {Alzheimer's}
  disease: overview and reproducible evaluation.
\newblock \emph{Medical Image Analysis}, page 101694, 2020.

\bibitem[Whitaker and Guest(2020)]{whitaker2020bropenscience}
K.~Whitaker and O.~Guest.
\newblock \#bropenscience is broken science.
\newblock \emph{The Psychologist}, 33:\penalty0 34--37, 2020.

\bibitem[Wilkinson et~al.(2020)Wilkinson, Arnold, Murray, van Smeden, Carr,
  Sippy, de~Kamps, Beam, Konigorski, Lippert, et~al.]{wilkinson2020time}
J.~Wilkinson, K.~F. Arnold, E.~J. Murray, M.~van Smeden, K.~Carr, R.~Sippy,
  M.~de~Kamps, A.~Beam, S.~Konigorski, C.~Lippert, et~al.
\newblock Time to reality check the promises of machine learning-powered
  precision medicine.
\newblock \emph{The Lancet Digital Health}, 2020.

\bibitem[Willemink et~al.(2020)Willemink, Koszek, Hardell, Wu, Fleischmann,
  Harvey, Folio, Summers, Rubin, and Lungren]{willemink2020preparing}
M.~J. Willemink, W.~A. Koszek, C.~Hardell, J.~Wu, D.~Fleischmann, H.~Harvey,
  L.~R. Folio, R.~M. Summers, D.~L. Rubin, and M.~P. Lungren.
\newblock Preparing medical imaging data for machine learning.
\newblock \emph{Radiology}, page 192224, 2020.

\bibitem[Winkler et~al.(2019)Winkler, Fink, Toberer, Enk, Deinlein,
  Hofmann-Wellenhof, Thomas, Lallas, Blum, Stolz,
  et~al.]{winkler2019association}
J.~K. Winkler, C.~Fink, F.~Toberer, A.~Enk, T.~Deinlein, R.~Hofmann-Wellenhof,
  L.~Thomas, A.~Lallas, A.~Blum, W.~Stolz, et~al.
\newblock Association between surgical skin markings in dermoscopic images and
  diagnostic performance of a deep learning convolutional neural network for
  melanoma recognition.
\newblock \emph{JAMA Dermatology}, 155\penalty0 (10):\penalty0 1135--1141,
  2019.

\bibitem[Woo et~al.(2017)Woo, Chang, Lindquist, and Wager]{woo2017building}
C.-W. Woo, L.~J. Chang, M.~A. Lindquist, and T.~D. Wager.
\newblock Building better biomarkers: brain models in translational
  neuroimaging.
\newblock \emph{Nature Neuroscience}, 20\penalty0 (3):\penalty0 365, 2017.

\bibitem[Yu et~al.(2018)Yu, Zheng, Liu, Huang, and Ding]{yu2018classify}
X.~Yu, H.~Zheng, C.~Liu, Y.~Huang, and X.~Ding.
\newblock Classify epithelium-stroma in histopathological images based on deep
  transferable network.
\newblock \emph{Journal of Microscopy}, 271\penalty0 (2):\penalty0 164--173,
  2018.

\bibitem[Zech et~al.(2018)Zech, Badgeley, Liu, Costa, Titano, and
  Oermann]{zech2018variable}
J.~R. Zech, M.~A. Badgeley, M.~Liu, A.~B. Costa, J.~J. Titano, and E.~K.
  Oermann.
\newblock Variable generalization performance of a deep learning model to
  detect pneumonia in chest radiographs: a cross-sectional study.
\newblock \emph{PLoS Medicine}, 15\penalty0 (11):\penalty0 e1002683, 2018.

\bibitem[Zendel et~al.(2017)Zendel, Murschitz, Humenberger, and
  Herzner]{zendel2017good}
O.~Zendel, M.~Murschitz, M.~Humenberger, and W.~Herzner.
\newblock How good is my test data? introducing safety analysis for computer
  vision.
\newblock \emph{International Journal of Computer Vision}, 125\penalty0
  (1-3):\penalty0 95--109, 2017.

\bibitem[Zhou et~al.(2020)Zhou, Greenspan, Davatzikos, Duncan, van Ginneken,
  Madabhushi, Prince, Rueckert, and Summers]{zhou2020review}
S.~K. Zhou, H.~Greenspan, C.~Davatzikos, J.~S. Duncan, B.~van Ginneken,
  A.~Madabhushi, J.~L. Prince, D.~Rueckert, and R.~M. Summers.
\newblock A review of deep learning in medical imaging: Image traits,
  technology trends, case studies with progress highlights, and future
  promises.
\newblock \emph{Proceedings of the IEEE}, pages 1--19, 2020.

\end{thebibliography}

\appendix
\section{Supplementary materials}

\subsection{Literature popularity review methods}\label{sec:popularity}

We give here the methodological details behind
figure~\ref{fig:publications}.
To assess relative popuplarity of studies on breast versus lung cancer in 
medical and AI research, we quantify the prevalence of these topics in
the corresponding  
literatures. For this, we use the
Dimensions.AI app~\citep{mori2018dimensions}, querying the titles and
abstracts of papers, with the following two queries:
\begin{itemize}
    \item lung AND (tumor OR nodule) AND (scan OR image)
    \item breast AND (tumor OR nodule) AND (scan OR image)
\end{itemize}
We do this for two categories, which are the largest subcategories within top-level categories ``medical sciences'' and ``information computing'': 
\begin{itemize}
    \item 1112 Oncology and Carcinogenesis
    \item 0801 Artificial Intelligence and Image Processing
\end{itemize}
We then normalize the number of papers per year, by the total number of
papers for the ``cancer AND (scan OR image)'' query in the respective
categories (1112 Oncology or 0801 AI).

\subsection{Details on Kaggle challenges studied}\label{sec:challenges}

Table \ref{tab:kaggle} gives details on the Kaggle challenges that we use
to compare performance gains to evaluation noise
(\autoref{sec:evalerror}).

For each competition, we looked at the public and private leaderboards, extracting the following information:
\begin{itemize}
    \item Differences $d_i$, defined by the difference of the $i$-th algorithm between the public and private leaderboard
    \item Distribution of $d_i$'s per competition, its mean and standard deviation 
    \item The interval $t_{10}$, defined by the difference between the best algorithm, and the ``top 10\%'' algorithm
\end{itemize}

\begin{table*}
    \centering
    \begin{tabular}{p{5cm} p{6cm} l l l}
        Description & URL & Incentive & Test size & Entries \\
        \hline
        
         Lung cancer detection in CT scans &
\url{https://www.kaggle.com/c/data-science-bowl-2017} & 1M USD & max 1K & 394 \\
         
        Schizophrenia classification in MR scans &  \url{https://www.kaggle.com/c/mlsp-2014-mri/overview} & Publications & 120 & 313 \\
         
         Lung tumor segmentation in X-rays & \url{https://www.kaggle.com/c/siim-acr-pneumothorax-segmentation} & 30K USD & max 6K & 350 \\
         
         Nerve segmentation in ultrasound images & \url{https://www.kaggle.com/c/ultrasound-nerve-segmentation} & 100K USD & 5.5K & 922 \\
         
    \end{tabular}
    \caption{Details of Kaggle challenges used for our analysis. The test size shows the number of test images provided, and the number of entries corresponds to the number of results on the private leaderboard.}
    \label{tab:kaggle}
\end{table*}

\end{document}